\documentclass[12pt]{article}
\usepackage{authblk}    
\usepackage[a4paper,vmargin=2.54cm,hmargin=3.17cm,]{geometry}
\usepackage{amssymb}
\usepackage{latexsym}
\usepackage{amsmath}
\usepackage{amsfonts}
\usepackage{hyperref}
\usepackage{graphicx}
\usepackage{natbib}
\setcitestyle{square,numbers}	
\begin{document}
\title{\textbf{Quantum computing using continuous-time evolution}}
\author{Viv Kendon\thanks{viv.kendon@durham.ac.uk; ORCID: 0000-0002-6551-3056}}
\affil{Physics Department, Durham University, Durham DH1 3LE, UK}
\maketitle              
\begin{abstract}
Computational methods are the most effective tools we have besides scientific experiments to explore the properties of complex biological systems.  
Progress is slowing because digital silicon computers have reached their limits in terms of speed.
Other types of computation using radically different architectures, including neuromorphic and quantum, promise breakthroughs in both speed and efficiency.  
Quantum computing exploits the coherence and superposition properties of quantum systems to explore many possible computational paths in parallel.
This provides a fundamentally more efficient route to solving some types of computational problems, including several of relevance to biological simulations.
In particular, optimisation problems, both convex and non-convex, feature in many biological models, including protein folding and molecular dynamics.
Early quantum computers will be small, reminiscent of the early days of digital silicon computing.
Understanding how to exploit the first generation of quantum hardware is crucial for making progress in both biological simulation and the development of the next generations of quantum computers.
This review outlines the current state-of-the-art and future prospects for quantum computing, and provides some indications of how and where to apply it to speed up bottlenecks in biological simulation.
\end{abstract}
\section{Introduction}\label{sec:intro}
In the quest for more computing power, our dominant digital silicon architectures have reached the limit of physically practical processor speeds.  
The heat conduction of silicon limits how fast waste heat can be extracted, in turn limiting the processor speeds.
Moreoever, energy consumption by computers is now a significant fraction of humanity's energy use\footnote{%
Estimates are around 3\% of global electricity use in 2020, depending on what you include.},
raising sustainability questions in the current global warming situation.
Today's silicon devices are orders of magnitude away from optimal in their energy use, but improving efficiency requires radical redesign.
We can't afford to apply more and more standard computers to solve the biggest problems,
we need more energy-efficient computational materials, and more efficient ways to compute.
This is especially relevant for computational modelling of biological systems, where their complexity means analytical or phenomenological models are inadequate for the level of explanatory and predictive power we seek.

Quantum computing promises fundamentally more efficient computation, at least for some important types of problems, such as simulation of quantum systems 
\cite{Brow2010}, 
non-convex optimisation 
\cite{Dick2013}, 
and (famously) factoring large semi-primes 
\cite{Shor1994,Shor1995}.
Areas where quantum enhanced accuracy could make a significant improvement to biological simulations include
configuration dynamics, such as protein folding and molecular docking interactions, and
electronic force field calculation for biomolecular modelling.
The former is a non-convex optimisation problem, a complex energy landscape in which minimisation processes can become trapped in local minima.
Quantum mechanical tunnelling can escape such local minima more easily than classical stochastic processes.
The latter is a convex optimisation problem, but with a fermionic quantum mechanical wave function that is notoriously difficult for classical computers (the \emph{sign problem}).
By using suitable transformations (e.g., Jordan-Wigner, or Bravyi-Kitaev), quantum systems can account for the antisymmetric nature of the electron wave function intrinsically.

The dominance of general purpose digital silicon computers has led to a lack of awareness of the breadth of diverse ways to encode and process information.
Brains, bacteria 
\cite{Hors2017}, 
and analog computers 
\cite{Shan1941} 
all compute very differently from digital silicon.
Neuromorphic architectures 
\cite{neuromorphic} 
and other unconventional hardware
(e.g., reservoir computers 
\cite{Dale2018}) 
are also promising low power alternatives for suitable types of problems.

This survey focuses on the role of quantum computers that are likely to be available and effective in the near term.  
It took fifty years of engineering from the first commercial transistors to today's smart phones, and we are not going to see advanced quantum computers appearing overnight.
Nonetheless, the pace of development is rapid, and we can expect the first  useful devices within a few years.  
The challenge is how best to exploit them in combination with our state-of-the-art digital classical computers.
The review is organised as follows.
In section \ref{sec:qc}, the basic ideas of quantum computing are explained, and some of the various different models outlined.
Quantum simulation is especially important for biomolecular simulation, and is covered in section \ref{sec:qsim}.
The principles and some examples of hybrid quantum-classical computing are
introduced in section \ref{sec:hybrid}.
The different types of hardware being developed to build quantum computers are briefly covered in section \ref{sec:hardware}, to illuminate the timeline for the availability of useful quantum computers.
Finally, the outlook for future advances is discussed in section \ref{sec:summary}.

\section{Quantum computing}\label{sec:qc}
One of the hallmarks of computing is that the same computation can be carried out in a variety of ways using different physical systems
\cite{Hors2013}.
As a simple example, consider computing the sum of two plus three.  
You have probably already arrived at the answer -- five -- using your brain to do mental arithmetic.
You could also use the calculator on your smart phone, or shuffle the beads on an abacus, for example.
Brains, smart phones, and abacuses are three very different physical things, yet they can all be used to perform the same computation.

For more powerful computation, digital silicon technology currently dominates, but sixty years ago, silicon transistor-based computers were a new technology in a setting where most computation was done by human brains, and analog devices for solving differential equations
\cite{Shan1941}.
We have now reached the limit of how fast computers based on this dominant silicon technology can compute.  
Until about 2009, faster silicon chips were produced by shrinking the size of the transistors.
Smaller transistors mean the electrical signals have shorter distances to travel at the same speed, hence they can operate faster.
However, the waste heat produced during computation needs to be dissipated, or the chips would melt.
Smaller transistors means the same amount of heat is produced in a smaller volume, and we have reached the limit imposed by the heat conduction of silicon -- we can't remove the heat fast enough to gain any increase in speed by further shrinking the transistors.

Increasing computing power is now fuelled by running more processors in parallel, but this, too, has limits.
Large supercomputing facilities fill warehouses with racks of computers, and the power required to run them is a significant drain on human energy resources.
The good news on this aspect is that conventional silicon chips are very energy inefficient, so radical new designs have the potential to leverage more computing for the same power consumption.
One example of customised lower power silicon chips is the SpiNNaker project 
\cite{neuromorphic}.

There are still gains to be made by optimising the design of silicon processors to suit different types of computation.  
Chip makers, such as Intel, produce hundreds of types of computer chips, optimised for smart phones, notebooks, laptops, desktops, or servers, not to mention cars, toasters, satellites, and everything in between that now has a dedicated computer to control it.
This trend is not new, the dedicated processor on a graphics card to control the display on your computer monitor has been standard for decades.
As these GPUs (graphics processing units) became more powerful, they were also exploited for computation, and a well-appointed HPC (high performance computing) facility will generally have banks of GPUs available alongside more conventional CPUs (central processing units).

Computing is thus entering an age in which progress requires the exploration and development of different types of computational hardware beyond the dominant silicon semiconductors of the past half century.
There are many strands to the research into new types of computers, but one of the most established, and most promising in terms of the potential advantages, is quantum computing.
Quantum computing exploits two properties of quantum systems to access a fundamentally different logical operation from classical computing.
Quantum mechanics allows systems to be in a \emph{superposition} of two or more different states at the same time.  
And it also allows stronger correlations to exist between systems than are possible in classical mechanics. 
These quantum correlations are known as \emph{entanglement}.
The basic building block of a digital quantum computer is a quantum bit, or \emph{qubit}, the quantum equivalent of a classical bit, a binary digit that can take the value zero or one.
Unlike classical bits, which are definitely either zero or one, a qubit can be in a superposition of zero and one at the same time.
The superposition has an important property called \emph{coherence}.
Quantum mechanics is a wave mechanics, like water waves or vibrating strings.
When two waves arrive at the same place, if they are in phase with each other, they add up to make a bigger wave, but if they are out of phase they cancel, leaving a smaller wave, or none at all.
A qubit in a superposition of zero and one has a specific phase between the zero and one components.   
This is critical for quantum computing, if the phase gets smeared out, the qubit behaves more like a classical bit, and is said to have \emph{decohered}.
A good quantum computer needs to keep its qubits coherent throughout the computation.
A more detailed introduction to quantum computing at a non-expert level can be found in
\cite{Kend2017}.

\begin{figure}
  \begin{center}
    \includegraphics[width=0.6\columnwidth]{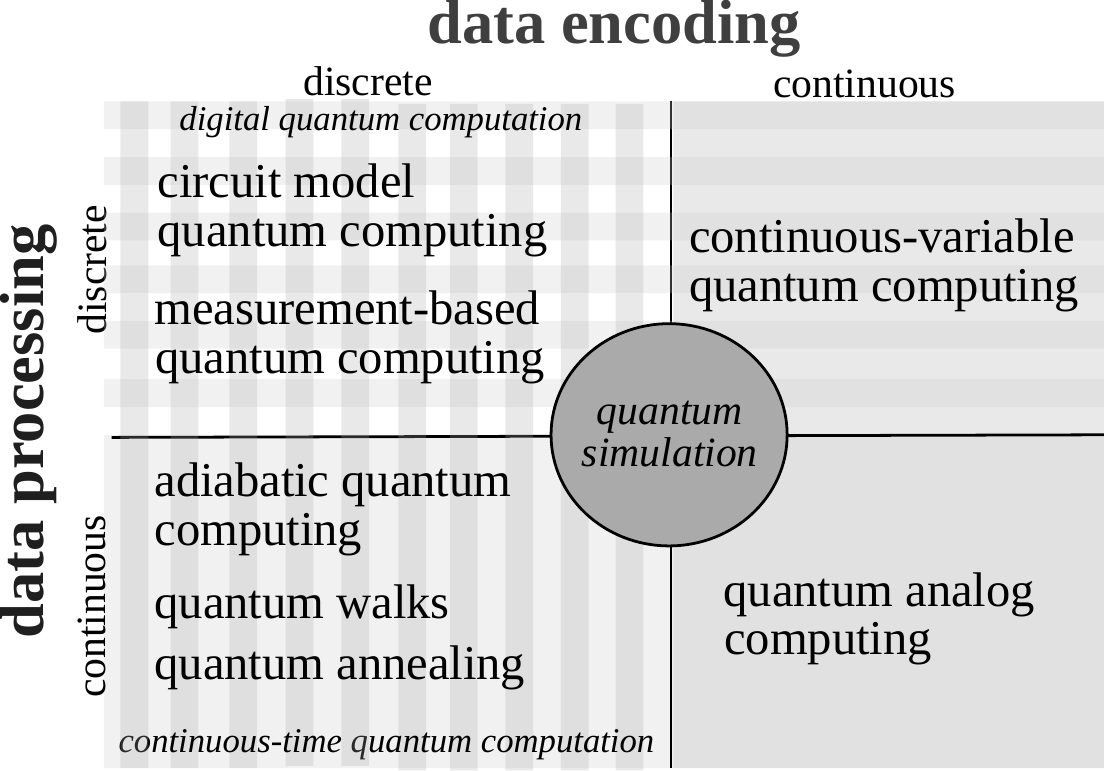}
  \end{center}
  \caption{%
    Diagram illustrating how quantum computing models use either discrete or continuous data encoding and processing.  The various models are described in the text.
    \label{fig:compmodels}
  }
\end{figure}

In keeping with the idea that computing needs to diversify, quantum computing itself has several different models under active development.  
Figure \ref{fig:compmodels} shows how some of the models are related, and the most relevant
are briefly outlined next. 
More varieties are covered in
\cite{Kend2017}.

\subsection{Circuit model quantum computing}\label{ssec:circuit}
\begin{figure}
  \begin{center}
    \includegraphics[width=0.7\columnwidth]{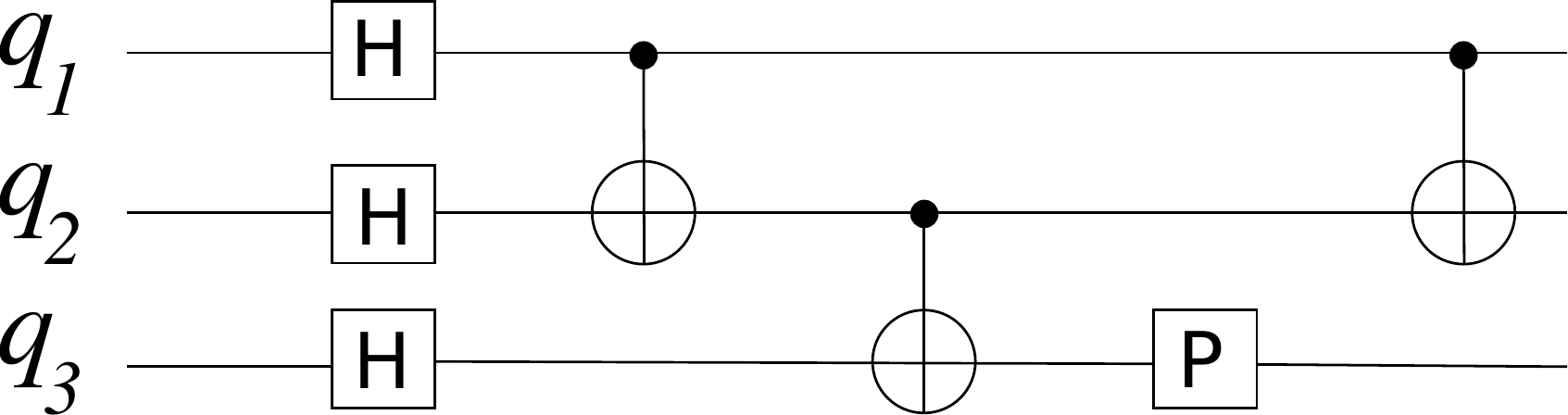}
  \end{center}
  \caption{%
    A simple circuit model quantum computing diagram. 
    Each horizontal line represents a qubit, labelled $q_1$ to $q_3$ on the left.
    The boxes with letters represent single qubit gate operations, \textsf{H} for a Hadamard gate and \textsf{P} for a phase gate.
    The circle connected to a dot represents a two-qubit controlled-NOT gate operation in which the state of the qubit under the dot determines whether the qubit under the circle is flipped.
    Time runs from left to right showing the order of the gate operations.
    This is just an illustrative fragment, actual computations would have many more gate operations in total.
    \label{fig:qcircuit}
  }
\end{figure}
The gate or circuit model is most closely analogous to digital classical computing. 
Quantum gate operations alter the state of a qubit in specific ways, for example, flipping the value from zero to one or one to zero, depending on what state it started in.  
If it started in a superposition of zero and one, both components are flipped.
Quantum gate operations can also act on two (or more) qubits, such that the state of one qubit is used to decide whether to update the other qubit, a controlled gate operation.
Bit flip and controlled bit flip are equivalent to classical gates (\textsc{not}, and controlled-\textsc{not}, respectively).
There are also qubit gate operations that do not have classical equivalents, such as the Hadamard and phase gates in figure \ref{fig:qcircuit}. 
These gate operations can change a qubit from a zero or one state into a superposition state.
To carry out a quantum computation, a register of qubits is initialised as all zeros, and a sequence of one and two qubit gate operations is applied to the qubits to perform the computation, see figure \ref{fig:qcircuit}.  
Finally, the qubit register is measured, the resulting sequence of zeros and ones is interpreted as the outcome of the computation.
Under the hood, your computer and smart phone work in much the same way using classical bits and classical gates in place of qubits and quantum gate operations.
For both classical and quantum, any possible computation can be built up using gates drawn from a small set of one and two (qu)bit gates, applied in the right order.
There are several such universal sets of gate operations, so it is possible to choose a convenient set to suit the hardware characteristics.

There are a few important differences to note in quantum computing.
It is a fundamental property of quantum mechanics that it is impossible to exactly clone an unknown quantum state 
\cite{Woot1982,Diek1982,Buze2001}.
This places restrictions on how quantum programs are written, and also makes it harder to correct for errors that may occur due to imperfect quantum hardware.
Classical computers are engineered to make the error rate very low, so low that you don't have to worry about it and can assume it works perfectly for most purposes.
This is not so easy to achieve in a quantum computer, where any interaction, including the gate operations necessary for performing the computation, can potentially cause an error.
Even qubits just left waiting until they are needed accumulate errors, due to the unavoidable interactions with their environment.
There is also a trade off between low error rates and the speed of gate operations that means that practical quantum computers unavoidably have significant error rates and will thus need some form of error correction.
The theory for this is well-developed
\cite{Devi2009,Lida2013}, 
but the number of extra qubits required is large, maybe a hundred or a thousand physical qubits to protect one logical computational qubit.
Consequently, we do not expect circuit model quantum computers to be the first useful quantum computational devices, unless there is a radical breakthrough in quantum error correction theory.

\subsection{Measurement-based quantum computing}\label{ssec:mbqc}
\begin{figure}
  \begin{center}
    \includegraphics[width=0.6\columnwidth]{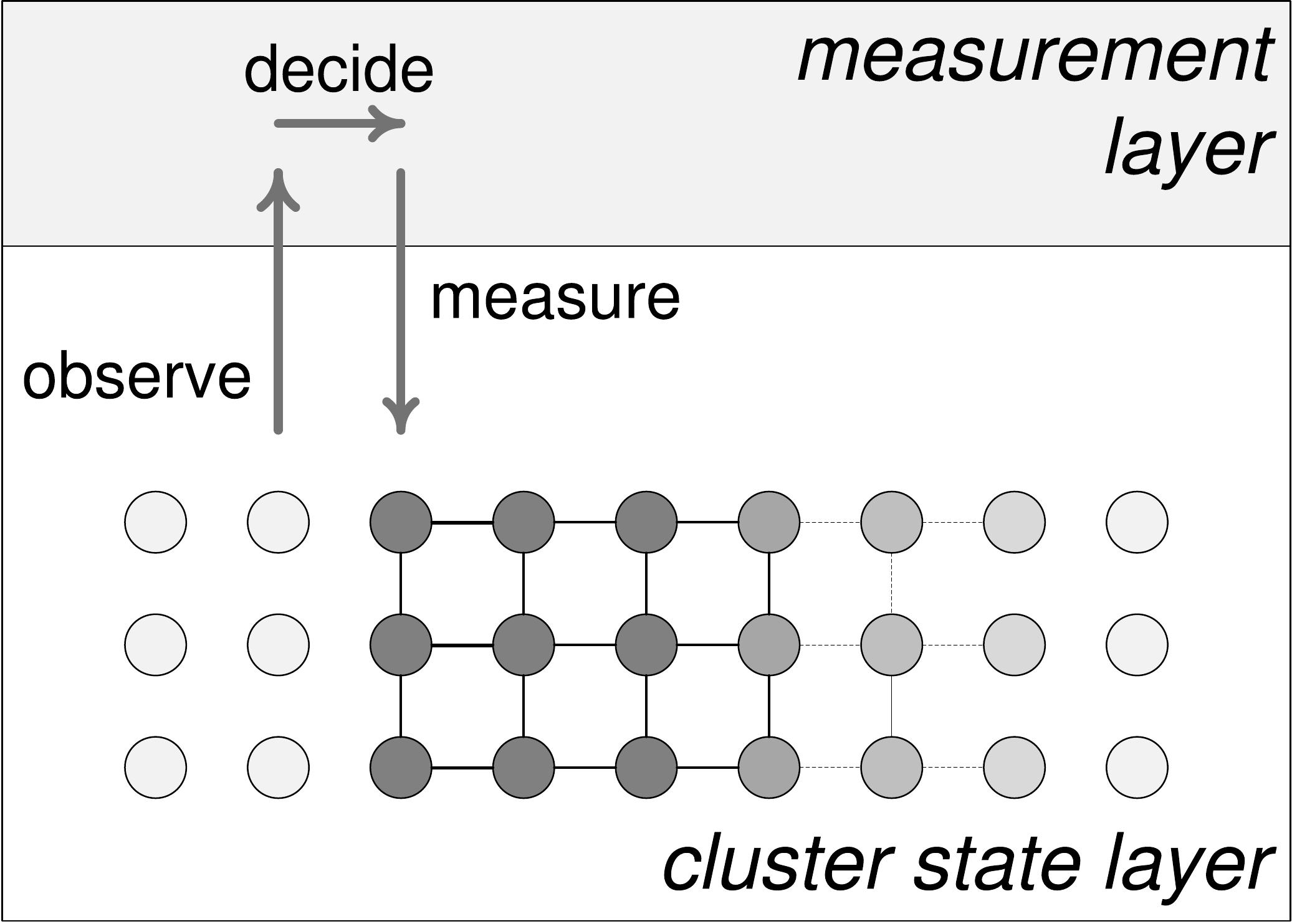}
  \end{center}
  \caption{%
    Measurement-based quantum computing diagram. The qubits (circles in the lower half) are first entangled with their neighbours (represented by lines) to form a cluster state.  Qubits on the right are being entangled to extend the cluster state.
    Measurements are made on each qubit in turn, which destroys the entanglement (qubits on the left have been measured).
    Later measurements can depend on the outcomes of earlier measurements: the top layer shows the classical control and decision-making process to determine the later measurements.
    \label{fig:mbqc}
  }
\end{figure}
In the early 2000s, Raussendorf and Briegel 
\cite{Raus2001} 
invented a model of quantum computing that first builds a
highly entangled state -- known as a cluster state -- of many qubits, then measures the qubits in turn to perform the computation, see figure \ref{fig:mbqc}.
The final set of measurements provide the outcome of the computation.
The earlier measurements are the equivalent of the gate operations in the circuit model.
The measurements are done with different phases, that can depend on the outcomes of previous measurements in the sequence, this is enough to generate any possible quantum computation, making it equivalent to the circuit model quantum computer.
It turns out that this approach can provide extra efficiencies beyond the circuit model, for example, to speed up Fourier transforms, at a cost of using more qubits.
Another advantage is that it is possible to build the entangled state just ahead of the measurements, so any given physical qubit only has to stay error-free for a fraction of the time taken for the whole computation 
\cite{Hors2010}.
This helps to reduce the errors that accumulate during a quantum computation, but does not eliminate them entirely.
Error correction can be combined with the measurements in much the same way as it is done for the circuit model.
Since measurement-based quantum computing requires more qubits than the circuit model, it is also not expected to be the design of the first useful quantum computers.
But there are potential advantages for scaling up to larger sizes, and the measurement outcomes provide a useful monitor of how the computation is proceeding.
Historically, it was also a useful reminder that quantum computing does not have to look like classical computing, and further distinct models of quantum computing were subsequently invented.

\subsection{Analog quantum computing}\label{ssec:analog}
Classical analog computing predates digital silicon computers.
In the first half of the last century, physics and engineering departments would often have an analog computer made from electronic components or water pipes or other ingenious mechanical constructions 
\cite{meccano}.
Analog computers solve differential equations, and their theory was put on firm foundations by Shannon 
\cite{Shan1941}.  The system is initialised, with the computation programmed into the settings of the resistors or valves or equivalent controls.  It is then allowed to evolve under the natural physics of the system (electrical circuits or hydrodynamics), and the state it reaches is measured to provide the solution. 

The main reason why digital computers displaced analog computers so easily is because of how the data are encoded.  
An analog computer represents numbers by the size of some physical quantity, a voltage or the height of a column of water, for example.
A digital computer represents numbers in binary, a string of zeros and ones, where a one in the $j$th position represents $2^j$, for example, nine is $101$ in binary.
The problem with analog is that if you want to double the precision of the numbers, you have to double the size of your computer.
Doubling the precision of a digital computer needs only one more bit, so eighteen is $1010$ in binary.
A single extra bit allows twice as many numbers to be represented.
Binary encoded numbers are a more efficient way to represent data.
Modern digital computers take efficient encoding to several more levels. Floating point numbers allow much larger or smaller numbers to be represented at a fixed precision, and data compression techniques allow redundant information (e.g., large patches of the same colour in an image) to be represented by one entry times the number of repeats.

Quantum analog computers are subject to the same drawback, i.e., the encoding of numbers is inefficient compared with digital quantum computers 
\cite{Blum2002}.
Like classical analog computing, the encoding of a number in a quantum analog computer is proportional to some quantity that can vary continuously (length, say, or frequency), rather than using qubits for a binary encoding. 
Nonetheless, a fully analog version of quantum computing is relevant for some quantum simulation and communications tasks.
A related version, known as \emph{continuous-variable} quantum computing
\cite{Brau2004},
in which the encoding is analog but the processing is carried out using discrete gate operations, is also potentially useful in the near future because it can be implemented using laser light, which leverages the highly developed laser technology we already have.
Analog quantum computing and continuous-variable quantum computing are able to carry out any possible quantum computation, but the resources required to encode the data will be inefficient for large computations.

\subsection{Continuous-time quantum computing}\label{ssec:ctqc}
There is another interesting option for quantum computers.  
Efficient binary encoding of data in qubits can be combined with the natural  evolution of physical systems continuously in time.
This combination leverages the fundamental difference between quantum mechanics and classical mechanics, as explained by Hardy 
\cite{Hard2001}.
Classical computing does not have an equivalent option for binary encoded data, because bits have only two possible values, zero and one.  
Changing from one to zero, or zero to one, is a discrete process (bit flip).
Qubits can change smoothly from zero to one through superpositions, so a continuous time evolution is both possible and natural.
Quantum mechanics is usually formulated as a Hamiltonian mechanics, i.e., there is a mathematical function, called the Hamiltonian, that describes how the law of energy conservation constrains the allowed changes.  
Starting from an initial quantum state, the Hamiltonian allows you to calculate how the state will change in future.  
For example, the Hamiltonian for a hydrogen atom (one electron, one proton) predicts the energy the electron can have while bound to the proton.  
As is typical for confined quantum systems, the possible energies are a discrete set.  
The electron can move to a different energy by absorbing or emitting a photon (quantum of light) carrying the extra energy.  
This gives rise to a set of observable spectral lines of wavelengths that are unique to hydrogen.  
Hydrogen atoms are not the most practical choice for building a quantum computer, but the basic idea, of navigating between different possible energies using photons, does carry over to practical designs.
We will return to this in section \ref{sec:hardware}.

The most common way to employ continuous-time quantum computing is to encode the problem to be solved in the Hamiltonian (energy function) applied to the qubits.  
Usually, the encoding is such that the answer is given by the ground state (lowest energy state) of the Hamiltonian.
The Hamiltonian can thus be viewed as a set of energy penalties, that make wrong answers correspond to higher energy states.
Efficient methods are known 
\cite{Choi2010,Luca2014}
to encode classical optimisation problems into suitable Hamiltonians, and much theoretical work has been done to characterise the computational properties of physically realisable Hamiltonians
\cite{Cubi2017}.
A suitable choice of Hamiltonian allows any possible quantum computation to be carried out, making continuous-time quantum computing equivalent to the quantum circuit model.
Within this model of continuous-time quantum computing, several distinct methods are known to enact the time evolution that carries out the computation.

\textbf{Adiabatic quantum computing} 
\cite{Farh2000} 
exploits the adiabatic theorem of quantum mechanics, which states that for systems with discrete energy levels, changing the Hamiltonian slowly enough will keep the system in the same energy level as it changes to match the Hamiltonian.
Starting in the ground state of a Hamiltonian that is easy to prepare, the Hamiltonian can be slowly transformed into a Hamiltonian with a ground state that encodes the answer to a hard problem.  
Measuring the qubits then provides the answer.  
The issue is then, how slowly is slow enough?  
For early quantum computers with limited quantum coherence, slow enough may not be possible.  
Interactions with the environment may disturb the process too much for the computation to be reliable.

\textbf{Quantum annealing} 
\cite{Fini1994}
is a quantum equivalent of simulated annealing, in which the system is cooled
into its ground state.  
For non-convex optimisation problems the computation can get stuck in a local minimum and fail to find the true global minimum solution.
Quantum systems have an extra trick to avoid getting stuck in a local minimum, separated from the true minimum energy by a high energy barrier: they can \emph{tunnel} through narrow barriers instead of having to jump over them.

\textbf{Continuous-time quantum walks} 
\cite{Farh1998}
are a quantum equivalent of classical random processes in which the system
evolves under an unchanging Hamiltonian that combines spin flips with terms that define the problem.  Childs
\cite{Chil2009}
proved that quantum walks are universal for quantum computing, provided they are encoded into a qubit register.
Their usefulness for hard optimisation problems is demonstrated in
\cite{Call2019}.

\begin{figure}
  \begin{center}
    \includegraphics[width=0.6\columnwidth]{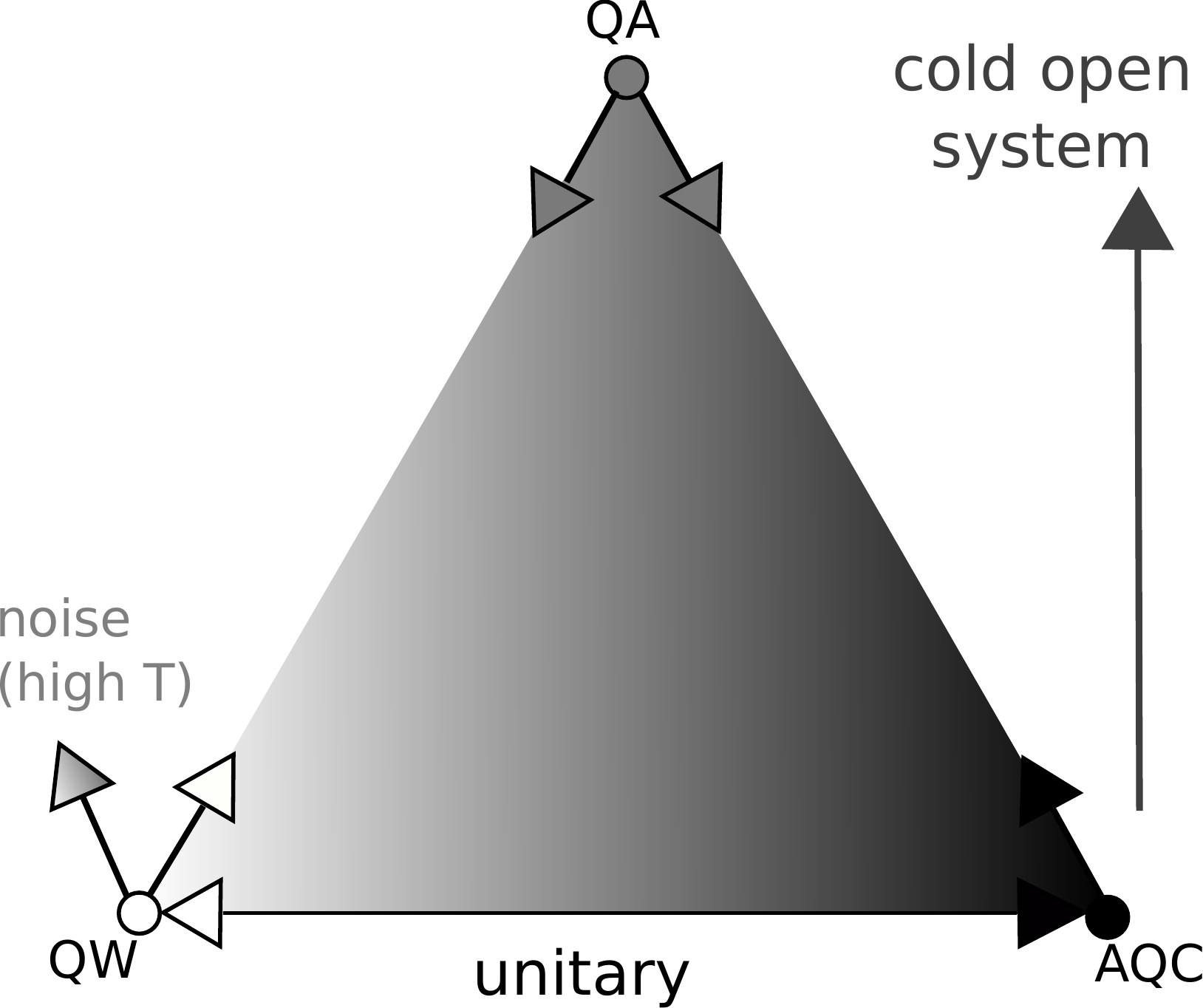}
  \end{center}
  \caption{%
    Continuous-time quantum computing diagram showing how quantum walk (QW), quantum annealing (QA) and adiabatic quantum computing (AQC) are related.  QW and AQC are pure quantum evolutions (unitary) while QA involves external cooling.  All can be subject to unwanted noise that causes errors.
    \label{fig:ctqctriangle}
  }
\end{figure}

The relationship between these three methods is illustrated in figure \ref{fig:ctqctriangle}.
Which method is best depends on the hardware, especially in a practical setting where quantum coherence may only persist for a short time, thus limiting the available time for the computation to run.  
Quantum walks and quantum annealing have shorter run times than adiabatic quantum computing, and hence may be preferable in practice. 
Hybrid algorithms combining the strengths of each method allow the best performance to be extracted from real hardware \cite{Morl2017}.

Continuous-time quantum computing is most often applied to solving classical optimisation problems.  
This is a subset of problems that doesn't require the full power of quantum computing, and can be engineered with simpler Hamiltonians than are required for fully universal quantum computing.
Continuous-time quantum computing for classical optimisation is especially promising for early quantum computing hardware.
The devices built by D-Wave Systems Inc.~have the largest number of qubits (around 2,000) currently available, although these are heavily influenced by their environment and lose coherence quickly relative to the time taken to compute.

In contrast to the circuit model and measurement-based quantum computing, for continuous-time quantum computing there is no well-developed theory of how error correction could be applied 
\cite{Youn2013}. 
Techniques from quantum control theory can help to mitigate errors, and smart methods for encoding problems also make the computations more robust against errors
\cite{Chil2002,Book2015}.
 Ultimately, there are limits to how large such quantum computers can be, because larger numbers of qubits require more precisely specified Hamiltonian parameters, and the controls to set the parameters have physical limits to their precision. 
In other words, there is as yet no theory for how to scale the hardware to arbitrary large sizes.
However, a few thousand qubits, even somewhat noisy ones, are still potentially a very powerful computational tool, so this is definitely an area to watch in the near term.

\section{Quantum simulation}\label{sec:qsim}

The original inspiration for quantum computing 
\cite{Feyn1982}
was to simulate quantum systems.
Quantum superpositions mean there are exponentially many more quantum states to keep track of in a simulation than for the equivalent size of classical system.
While there are many cases in which only a significantly smaller part of this huge state space is relevant, it still presents a challenge for numerical simulations.
Another basic quantum characteristic that makes numerical simulation difficult is indistinguishable identical particles.
Classical identical particles can always have an extra label attached to them to make them distinguishable.
With quantum particles this is not possible.  
We solve the problem for qubits by trapping each qubit in a different location, so that the chance of them spontaneously swapping places is negligible.
But in general, if two identical quantum particles can swap places, the physical description has to be invariant under that swap.
This permits just two types of quantum particle: Bosons and Fermions.
Bosons include photons (light), and many types of atomic nuclei with even numbers of nucleons (protons and neutrons).
Fermions include electrons, protons, and neutrons, the basic building blocks of matter.
Fermions are the awkward ones to simulate.
When they swap places, there is an extra quantum phase that appears in the equations, making them antisymmetric overall.  
This messes up Monte Carlo simulations and is known as the \emph{sign problem}.
The huge state space combined with the sign problem makes calculating the electron structure, one of the basic ingredients for molecular simulations, a formidable challenge.
This is one of the problems where we anticipate that quantum computers can soon be applied to do useful computation.
Larger and more accurate electronic structure calculations will be useful for biomolecular 
\cite{Harr2010,Hugg2018}
and materials simulations, with potential breakthrough applications from basic science, e.g., understanding enzyme catalysis, leading to smart materials and drug discovery.

Quantum simulation on general purpose digital quantum computers (circuit model) was shown to be  efficient compared to classical computers 
\cite{Lloy1996}, 
but calculating the equations of motion in discrete steps is still expensive even for a quantum computer.
A more promising approach in the near term is to build a quantum simulator that has the same Hamiltonian as the system being studied.
The quantum simulator doesn't need to be a universal quantum computer, it just needs to be able to simulate one particular Hamiltonian, or class of Hamiltonians.
Theoretical work has identified which Hamiltonians are capable of simulating which other Hamiltonians, guiding the design of quantum simulators 
\cite{Cubi2017}.
Calculating the equations of motion then simplifies to letting the quantum simulator evolve naturally for the desired period of time.
Thus, this type of special purpose quantum simulator is another example of continuous-time quantum computing.
Such quantum simulators can potentially be used to solve other types of problems that can be naturally encoded into their Hamiltonians, although this has not yet been explored in detail.
Systems capable of quantum simulation have been developed to an advanced stage, e.g., the quantum gas microscope
\cite{Kuhr2016},
but will require more engineering to make them suitable for operation beyond scientific laboratories.
Spin out companies and the UK's National Quantum Computing Centre
\cite{NQCC2020} are gearing up to tackle this technology translation process.

There are a wide variety of quantum simulation algorithms to match the wide range of problems that are interesting to study 
\cite{Brow2010}.
The most common approach is to prepare a quantum state that is a superposition of the ground state and first few excited states of the quantum Hamiltonian.
Then, the energy levels are calculated using phase estimation.
Phase estimation is used as a subroutine in many quantum algorithms.
It uses a Fourier transform to convert quantum phases into measurable data in the qubit register.
However, it is not a one shot calculation, the computation must be repeated enough times to obtain the desired accuracy in the phase being estimated.
In fact, repetition is a relatively common feature of quantum computations, because the outcome is not necessarily deterministic.  
Obtaining the correct outcome with a probability significantly better than guessing is often enough to gain an advantage.  
With enough repeats, the correct outcome can be identified with a high level of confidence.

\section{Combining classical and quantum computing}\label{sec:hybrid}

The first useful quantum computers will be relatively small, and limited in what they can do.
They are not going to sweep aside our impressively advanced digital silicon
computers.  
Rather, they will augment them, by speeding up bottlenecks that are hard for classical computers. 
Using co-processors for specialised tasks is well-established.  
Graphics cards have been standard for decades and dedicated chips for ethernet or wireless connections are common.
As well as normal computers and GPUs, field programmable gate arrays (FPGAs) and application specific integrated circuits (ASICs) are often now available in HPC facilities to speed up bottleneck subroutines.
Adding a quantum co-processor is a natural extension of this trend.

The diversifying range of computational devices becoming available presents several new challenges.
At the physical level, interfaces to allow the different types of hardware to communicate and pass data between them have to be implemented.  
This requires both the physical format of the signals to be matched, and the timing of the different types of processors to be synchronised.
Taking the example of a superconducting quantum computer, the electrical or optical signals from a standard computer at room temperature have to be converted into microwave signals at a temperature very close to absolute zero.
This introduces delays as the electronics perform the conversion.
The natural processing speeds of each device are different, although in this case they are both of the order of a few GHz (GigaHerz, $10^9$ operations per second).
Qubits have gate operation times set by the energy difference between the zero state and the one state.
In superconducting qubits this is matched to microwave frequencies in the GHz regime.
However, measurements often take longer than gate operations, and it is the measurements that provide the data sent back to the standard computer.

The theory of how to program combinations of different types of computational devices has not kept pace with practice, and it is something of an experimental art to obtain useful computing from such combinations.  
The first step is to identify which parts of the algorithm are most efficiently processed on which types of processor.
Some steps are intrinsically parallel and can be handled fastest by GPUs using shared memory.
Other steps may require the state of the whole simulation to be checked before further processing can proceed, and are best handled by a CPU.
The overhead of transferring the data between different processors also has to be taken into account.
To investigate whether a quantum co-processor could speed up an existing algorithm, the general method is to identify the most time-consuming elements in the algorithm -- the bottlenecks -- and find quantum algorithms that can potentially speed up these parts of the calculation.
Some examples of how to do this at the algorithmic level for quantum computing have been presented by Chancellor 
\cite{Chan2016,Chan2017}.

While there are general methods for different types of computation, each algorithm is specific to the application, and will need to be assessed individually for possible quantum enhancement.
This is the area in which work most needs to be done, in collaboration with user groups who know the algorithmic bottlenecks they face that quantum computers can potentially bypass 
\cite{Harr2010,CCPs}.
This will enable users to take advantage of quantum co-processors as soon as they become available alongside conventional HPC facilities.  
Test bed size quantum computers are becoming available to end users now, ready for early adopters to develop quantum enhancements to their algorithms.

\section{Quantum hardware}\label{sec:hardware}

Current quantum computing hardware comes in a variety of types, both to match the variety of theoretical models described earlier, and because there is not yet one dominant technology analogous to how silicon semiconductors dominate classical computing today.

\subsection{Circuit model quantum hardware}\label{ssec:circuithw}

It is possible to trap individual atoms or ions far enough apart to manipulate them using lasers or applied electric and magnetic fields.
One very effective way to do this uses laser light to make the traps, similar to the optical tweezers sometimes used to manipulate individual cells.
Trapped ions are the most advanced in terms of having the lowest error rates for the gate operations.
The qubit is identified with two convenient energy levels in the ion, with calcium ions being a common choice.
The ions are controlled by laser pulses of the right frequency and duration.
These can be arranged as small modules containing seven or so ions, with light-based connections between the modules
\cite{Nick2014}.
While the individual elements of this design have been demonstrated, they have yet to be assembled into a working whole.
Much engineering work is required to miniaturise the traps that hold the ions, and the laser systems to control them.
Note that the intrinsic length scale in such systems is set by the wavelength of the light used to make the traps.  
For visible light, this is hundreds of nanometres, much larger than the feature size (ten to twenty nanometres) in current silicon chips.
This will set a practical limit on the number of qubits in ion trap quantum computers that is lower than current silicon chips.  
While this is not a problem in the near future, it helps to explain why it is important to keep options open on different types of hardware.

Superconducting systems provide qubits as small loops of current flowing without resistance at the low temperature of operation, close to absolute zero. 
These can be controlled by magnetic fields, and by microwave fields produced by small resonators that couple adjacent qubits.
Google and IBM are among the commercial developers of superconducting qubit devices, and both currently claim to have about fifty working qubits on a single device, typically on a chip about four centimetres square.
These qubits and resonators are just visible to the naked eye.
Compared with the development of classical computers, bits were still of a size visible to the naked eye in the mid-seventies, as magnetic core memory.
These superconducting qubits have higher error rates than trapped ions, and while they can run test bed quantum algorithms that are challenging to simulate on a classical computer 
\cite{Goog2019}, 
they are not yet able to perform useful quantum computations.

As well as providing connections to and between qubits made of ions, atoms or superconducting circuits, light in the form of photons can also be used to make qubits.
Advantages include less need for cooling to extremely low temperatures, and easy interfacing with quantum communications for distributed computing applications. 
However, it is harder to make photons interact with each other, so controlled gate operations may need to be repeated many times until they succeed.
This is less of a problem for measurement-based quantum computing, where the entangled state can be built ahead of time, with allowances made for the repetitions that will be required.
Thus, photonic quantum computing developers tend to prefer measurement-based quantum computing over the quantum circuit model.
All-photonic quantum computing is being developed by several key players in academia and industry, and we might expect to see interesting test bed implementations in three to five years.

At an earlier stage of development are a variety of silicon-based designs.
Google and IBM's superconducting qubits are large mainly because this reduces the errors induced by the substrate.
Making tiny qubits in silicon is challenging to engineer; even more challenging is controlling the errors produced by all the extra silicon atoms around the qubits.
Using isotopically pure silicon of the form with no nuclear spin reduces this noise to much lower levels.  
There is not yet a ready source of sufficiently pure silicon, so the development of the qubit designs is continuing in natural silicon, with the knowledge that the performance can be stepped up significantly once the designs are perfected.
There are several different ways to make qubits in silicon, including implanting phosphorus atoms into the silicon crystal structure, one for each qubit
\cite{He2019}, 
and making single electron quantum dots
\cite{Zhao2019}.
Engineering the controls for the qubits is also very challenging, especially since they need to operate at very low temperatures.  
While one might expect that our highly advanced classical silicon computer technology can provide controls, remember that it produces heat, and has been developed to operate at room temperature.
Silicon semiconductor properties change at low temperatures.
These types of engineering challenges appear at every stage of quantum computing development.
There is no indication that we cannot overcome them, but they will take time to solve, just as today's smart phones did not appear overnight after transistors were developed.

\subsection{Continuous-time quantum hardware}\label{ssec:ctqchw}

In terms of the number of qubits, the largest examples of continuous-time quantum computing hardware are the devices made by D-Wave Systems Inc.
\cite{Dick2013}, 
which pack around 2000 qubits onto a four centimetre square chip.
The penalty for making superconducting qubits this small with today's state-of-the-art technology is that the qubits are severely affected by the surrounding atoms, and lose quantum coherence before the end of the computation.
The current design of these devices is a special purpose quantum annealer, which is not capable of generating a fully universal quantum Hamiltonian.
However, it is universal for classical problems that can be encoded into Ising Hamiltonians, and this includes many practical hard problems such as network routing, scheduling, stock control, supply chains, and other optimisation problems with applications across industry and beyond.
It is not clear whether 2000 imperfect qubits is enough to really challenge the capabilities of classical computers for these problems, but the next generation of their devices will have more qubits, and lower noise rates.

Also in the continuous-time category are the special purpose quantum simulators described in section \ref{sec:qsim}.
Besides the quantum gas microscope 
\cite{Kuhr2016}, 
there are large linear arrays of trapped ions
\cite{Bern2017},
trapped atoms, especially with electrons excited to high energy levels, known as Rydberg states
\cite{Siba2018}, 
and cold molecules trapped in arrangements from one dimensional lines to complex three dimensional structures
\cite{Mose2017}. 
These all have strong, long range interactions between the atoms or molecules, with various controls to allow specific many-body quantum Hamiltonians to be engineered.
While these systems are still confined to experimental labs, they are coming close to the stage where they can perform interesting simulations.

This is only a selection of the quantum devices under development, and it is too early to pick a clear winner among the pack.  
Most likely there will be a role for several of them at some stage in the development of quantum computers.
No one system appears to combine all the advantages as well as being scalable to very large numbers of qubits.
The best design for a small quantum computer that works well within five years is likely not on a direct path to the ultimate design for devices coming online fifty years from today.

\section{Outlook}\label{sec:summary}

Classical computers run out of power to simulate quantum systems at a size equivalent to around 50 qubits, and our silicon semiconductor technology has reached its speed limit.
New computational devices are needed to progress beyond current state-of-the-art, and quantum computers promise fundamentally faster computation by harnessing the quantum properties of materials.
To achieve this potential, there are many challenges to be overcome, and it is unrealistic to expect too much too soon.
Nonetheless, there are opportunities in the next few years for pioneers of computational science to make breakthroughs in our most challenging computational tasks.

Simulation of quantum systems is expected to be among the first useful applications, and there are opportunities for speeding up bottlenecks in other applications, especially where optimisation is required.
Since quantum simulation starts to become useful at around 50 qubits, there is the potential for useful and interesting computation on the somewhat noisy, imperfect qubits currently available, if configured for use with a continuous-time evolution.
Digital quantum computers configured for circuit model gate operations are not likely to be so useful until they have sufficient error correction, at sizes of thousands rather than hundreds of qubits.
Continuous-time quantum computers can thus bridge the gap until digital quantum computing is ready for large-scale deployment.
Early quantum computers for quantum simulation and optimisation also may not be capable of the full range of quantum computations.
This can simplify the engineering, providing a stepping stone to fully universal quantum computing hardware.
Quantum simulation using trapped atoms or ions, for example, may only be able to compute a limited range of quantum Hamiltonians.
And quantum annealers based on Ising Hamiltonians can only encode classical optimisation problems.

Nonetheless, the opportunities for computational biology to exploit near term quantum computers are significant, with atomistic biomolecular simulations that need electronic structure calculations among the first applications to consider for a quantum upgrade.  
Processing X-ray diffraction data for macromolecular crystalography is an example of an optimisation task that could be suitable for quantum algorithms, with wide-ranging applications in biology and medicine
\cite{From2015}, especially given the data rates involved in real time visualisation of electron dynamics.
Another workhorse of quantum algorithms is the quantum Fourier transform, which can provide a significant speed up when applied to data stored in a quantum superposition.
Note that the quantum advantage is only gained when embedded in a longer quantum algorithm
\cite{Brow2007},
which points to the importance of working with experts in quantum algorithms for designing quantum subroutines.

To support the development of suitable algorithms and proof-of-principle implementations, a Collaborative Computational Project in Quantum Computing has been set up
\cite{CCPs}
to network between quantum computing experts and application experts across a broad range of computational science.
The emerging model of diverse combinations of computing hardware is a practical way to enhance the available computing power, but is also very challenging to program effectively.
An interdisciplinary approach is essential in the pioneering stages of developing quantum enhancements for useful computational applications.
With the current investment by the UK in research and innovation funding for quantum technology topping \pounds 1 billion, the field is developing rapidly, and this is an excellent time to start developing new computational methods for hard problems in the life sciences.

\vspace{1em}

\noindent 
This work was supported by the UKRI Engineering and Physical Science Council fellowship grant number EP/L022303/1.


\end{document}